\documentstyle[12pt,epsfig]{article}

\def\mboxsc#1{\mbox{\scriptsize #1}}
\def\smin{s_{\mbox{\scriptsize min}}}

\def\sminn{s_{\mbox{\scriptsize min}}^{\mbox{\scriptsize nlo}}}

\catcode`@=11
\newcount\@tempcntc
\def\@citex[#1]#2{\if@filesw\immediate\write\@auxout{\string\citation{#2}}\fi
  \@tempcnta\z@\@tempcntb\m@ne\def\@citea{}\@cite{\@for\@citeb:=#2\do
    {\@ifundefined
       {b@\@citeb}{\@citeo\@tempcntb\m@ne\@citea\def\@citea{,}{\bf ?}\@warning
       {Citation `\@citeb' on page \thepage \space undefined}}%
    {\setbox\z@\hbox{\global\@tempcntc0\csname b@\@citeb\endcsname\relax}%
     \ifnum\@tempcntc=\z@ \@citeo\@tempcntb\m@ne
       \@citea\def\@citea{,}\hbox{\csname b@\@citeb\endcsname}%
     \else
      \advance\@tempcntb\@ne
      \ifnum\@tempcntb=\@tempcntc
      \else\advance\@tempcntb\m@ne\@citeo
      \@tempcnta\@tempcntc\@tempcntb\@tempcntc\fi\fi}}\@citeo}{#1}}
\def\@citeo{\ifnum\@tempcnta>\@tempcntb\else\@citea\def\@citea{,}%
  \ifnum\@tempcnta=\@tempcntb\the\@tempcnta\else
   {\advance\@tempcnta\@ne\ifnum\@tempcnta=\@tempcntb \else \def\@citea{--}\fi
    \advance\@tempcnta\m@ne\the\@tempcnta\@citea\the\@tempcntb}\fi\fi}
\catcode`@=12

\begin{document}
\title{\vskip-3cm{\baselineskip14pt
\centerline{\normalsize MPI/PhT/2001-09 \hfill}}
\vskip1.5cm
Combining Parton Showers with \\Next-to-Leading Order QCD 
Matrix Elements \\in Deep-Inelastic $eP$ Scattering}
\author{{\sc B. P\"otter, T. Sch\"orner}\\
{\normalsize Max-Planck-Institut f\"ur Physik
(Werner-Heisenberg-Institut),}\\
{\normalsize F\"ohringer Ring 6, 80805 Munich, Germany}}

\date{\today}

\maketitle

\thispagestyle{empty}

\begin{abstract}
We have implemented a systematic procedure for combining parton shower 
algorithms with next-to-leading order QCD calculations for the 
case of jet production in deep-inelastic electron-proton scattering. 
Using this method we have computed
inclusive jet cross sections and jet shapes for the case of single-jet
production and compared them to data from the ZEUS collaboration
at HERA. We found good agreement between the data and our
calculations, both for the jet shapes and the inclusive spectra.

\bigskip

\end{abstract}


\section{Introduction}

In the last two decades sophisticated event generators have been
developed for all relevant physical scattering processes. For the case
of $eP$ scattering, generators like {\tt PYTHIA} \cite{pythia}, 
{\tt HERWIG} \cite{herwig}, {\tt LEPTO} \cite{lepto} or {\tt RAPGAP}
\cite{rapgap} are available. In the simplest case the event generation
procedure falls into three steps. First, a one-parton final state is
generated from the ${\cal O}(\alpha_s^0)$ quark-parton-model (QPM)
matrix element. Next, a parton shower (PS) algorithm \cite{herwig,PS} is
attached to the one-parton final state. The PS takes into account
arbitrarily high orders in $\alpha_s$, but only in the leading 
$\log Q^2$ approximation as opposed to the exact treatment in fixed
order matrix elements. This implies that the region of small $k_\perp$
around the original QPM quark is described quite well, which is
important for the description of the internal properties of jets.
Finally the PS is terminated at some lower cutoff $Q_0^2$, where
hadronization models such as the Lund String model \cite{lund} can be
used to simulate the transition of partons to hadrons. 

The approximations involved in the construction of the PS allow to
describe well only the region of small angle parton emission. The
region of wide angle scattering should be more appropriately described
by the hard QCD matrix elements of ${\cal O}(\alpha_s)$ which describe
the QCD-Compton (QCDC) and boson-gluon fusion (BGF) processes with two
final state partons. To include these processes into the event
generators, the phase-space is separated into two complementary
regions by means of some technical separation parameter, which we call
$R_{tech}$ in the following. In one ('soft') region, the QPM result
plus PS is used to produce the cross section, whereas in the other
('hard') region the QCDC and BGF matrix elements are employed. The
approximations involved in the PS approach suggest that the separation
parameter $R_{tech}$ should not be chosen too large in order to
guarantee that wide angle scattering is  appropriately described by
the hard QCD matrix elements. One has to ensure that none of the
partons generated by the PS populates the hard region above $R_{tech}$,
since this would lead to double counting. Apart from separating the
phase space into complementary regions, the parameter $R_{tech}$
serves as a cutoff and ensures the finiteness  of the QCD matrix
elements, which are divergent in the region of soft and collinear
particle emissions. A problem occuring at this stage is that the
overall normalization of the cross section is not yet fixed. For the
special case we are discussing here, the total cross section can be
calculated beforehand, which can be used to normalize the
contributions coming from the QPM and QCD matrix elements. A lower
boundary for $R_{tech}$ is obtained by the  requirement that the sum
of the QCDC and BGF contributions does not exceed the total cross
section. Otherwise $R_{tech}$ can be chosen freely. 

The normalization of the individual contributions by calculating the
total cross section beforehand is in
general not feasible e.g.\ for the case of dijet production in $eP$
scattering.  It is therefore desirable to find a general, systematic
approach to fix the cross section when fixed order matrix elements are
combined with PS algorithms. To reliably calculate the normalization of
the cross section, the complete next-to-leading order (NLO)
corrections to the leading order (LO) processes have to be
calculated. This includes the calculation of the real soft and
collinear regions, as well as of the virtual corrections. The NLO
corrections to the QPM graph in deep-inelastic scattering (DIS) have
long been evaluated \cite{1jet}. Also for the DIS dijet case, several
NLO calculations are avaible \cite{disent,graudenz,mepjet,jetvip},
which employ either the subtraction method \cite{disent,ERT,KS} or the  
phase-space slicing technique \cite{graudenz,ps1,ps2,ps3} to treat the
soft and collinear part of the real corrections. The problem of
directly combining the NLO cross sections with the PS is the occurance
of large positive and negative weights in the NLO calculation which
makes it practically difficult to obtain  numerically stable
results. Furthermore, it is not straight-forward to use the
hadronization models with negative weights.

Recently, several proposals where made to combine NLO QCD
calculations, including the virtual corrections, with PS's \cite{1,2,3,4,5}.
In this paper, we will rely on the method proposed by one of us
(B.P.) in \cite{4} and apply it to the case of single-jet inclusive
cross sections in DIS $eP$ scattering. The idea is to keep the
separation of the phase-space into a soft and a hard region with help
of the $R_{tech}$ parameter, but to use the full NLO
calculation in the soft region instead of the LO one. To ensure that the
weights generated in the soft region are always positive, a method
adapted from \cite{oldpc} is employed where the $\smin$ parameter of
the phase-space slicing method is chosen such that the sum of the
Born, virtual and soft and collinear contributions vanish. In this
way, the NLO corrections are calculated from the hard matrix elements,
integrated within the soft region down to the $\smin$ cutoff, which will
always yield positive weights. While in \cite{oldpc} the cutoff was
adjusted by hand, in \cite{4} it is calculated from the NLO
corrections, thereby preserving the improved scale and scheme
dependence of the NLO calculation.

The outline of the paper is as follows. In section 2 we specify how to
generate events with positive NLO weights and how we have implemented
the method for the case of single-jet production in DIS $eP$
scattering. In the following section we calculate inclusive jet cross
sections and jet shapes as measured by the ZEUS collaboration at HERA
for large $Q^2$ in the laboratory frame. We conclude with a short
summary and an outlook.

\section{Event Generation with Positive NLO Weights}

In $eP$ scattering
\begin{equation}
  e(k) + P(p) \to e(k^\prime) + X
\end{equation}
the simplest hadronic final state consists of a single jet with a large
transverse energy $E_T$ in the laboratory frame. The lowest order
${\cal O}(\alpha_s^0)$ partonic contribution to this single-jet cross section
arises from the QPM subprocess. At NLO, the single-jet cross section
receives contributions from the real and the one-loop virtual
corrections. The real corrections consist of the BGF and QCDC
processes which are divergent for collinear and soft emissions. These
divergencies are cancelled by the virtual corrections or are absorbed
into the parton density functions (PDFs) of the proton. To enable a
numerical treatment of the real corrections, one can employ
the phase-space slicing technique and introduce a parameter $\smin$ which
cuts out the singular regions. Within this method, the NLO cross section
can be written as a sum of the one-parton final state up to ${\cal
O}(\alpha_s)$ and the two-parton final state. The one-parton final
state reads
\begin{eqnarray}
 \sigma^{\mboxsc{1parton}}_{\mboxsc{had}}(\smin) 
 &=& \sigma_0 \
 \sum_{i=q,\bar{q}} e_i^2\ \int dx \ d{\mbox{PS}}^{(k^\prime+1)} \ 
 \bigg[ f_i(x,\mu_F) \left(1 + \alpha_s(\mu_R)\ {\cal{K}}_{q\rightarrow
 q}(\smin,Q^2) \right) \nonumber \\
 &+& \alpha_s(\mu_R)\ C_i^{\overline{\mboxsc{MS}}}(x,\mu_F,\smin) \bigg]
 |M_{q\rightarrow q}|^2  \label{1parton} \ .
\end{eqnarray}
The $f_i(x,\mu_F)$ are the proton PDFs, the Lorentz-invariant phase
space measure $d\mbox{PS}^{(k^\prime+n)}$ contains both the
scattered electron and the partons from the photon-parton scattering
process and the term $\sigma_0$ is defined as
$\sigma_0=4(\pi\alpha)^2/(Q^4xs)$. The $e_i$ are the charges of the 
quarks. The factor ${\cal{K}}_{q\rightarrow q}$ which depends both  
on the phase-space slicing parameter $\smin$ and on the hard scale of the
process $Q^2$ contains the virtual and final state corrections and is
specified in \cite{4}. The function $C_i^{\overline{\mboxsc{MS}}}$
containing the initial state corrections is given by 
\begin{equation}
C_i^{\overline{\mboxsc{MS}}}(x,\mu_F,\smin)=
\left(\frac{N_C}{2\pi}\right)
\left[ A_i(x,\mu_F)\ln\left(\frac{\smin}{\mu_F^2}\right)
+      B_i^{\overline{\mboxsc{MS}}}(x,\mu_F)\right] \ .
\end{equation}
The functions $A_i(x,\mu_F)$ and
$B_i^{\overline{\mboxsc{MS}}}(x,\mu_F)$ are also specified in
\cite{4}. The final result, independent of $\smin$, is obtained by
adding to the one-parton contribution the contribution containing the
two-parton final state, integrated over those phase-space regions
where any pair of partons $i,j$ has $s_{ij}>\smin$ with
$s_{ij}=(p_i+p_j)^2$: 
\begin{eqnarray}
 \sigma^{\mboxsc{2parton}}_{\mboxsc{had}}(\smin) &=& \sigma_0 \
 \sum_{i=q,\bar{q}} e_i^2 \!
 \int\limits_{|s_{ij}|>\smin} \!\!\!\!\!\!  dx \ d\mbox{PS}^{(k^\prime+2)}
 \ 4\pi\alpha_s(\mu_R)\ \bigg[ f_i(x,\mu_F)] \ |M_{q\rightarrow qg}|^2
 \nonumber \\ &+&  \mbox{$\frac{1}{2}$} \, f_g(x,\mu_F) \ 
 |M_{g\rightarrow q\bar{q}}|^2 \bigg] \ .  \label{2parton}
\end{eqnarray}

The condition 
\begin{equation}
\frac{d\sigma^{\mboxsc{1parton}}_{\mboxsc{had}}}{dx\,dQ^2}(\sminn)=0
\end{equation}
can be used to determine the function \cite{4}
\begin{equation}
 \sminn(\mu_F,\mu_R,x,Q^2) = \exp \left[ \eta - \sqrt
  { \eta^2 + \psi } \,\, \right]   \label{master}
\end{equation}
in which
\begin{eqnarray}
  \eta &=& \ln(Q^2) - \frac34 + \frac{9}{16} \frac{A}{F} \ , \\ 
  \psi &=&  -\ln^2(Q^2) + \frac{3}{2}\ln(Q^2)
  -\frac{\pi^2}{3}-\frac{1}{2} + \frac98 \left[ \frac{2\pi}{N_C\alpha_s} +
  \frac{B}{F} - \frac{A}{F} \ln(\mu_F^2) \right] 
\end{eqnarray}
and 
\begin{eqnarray}
 F &=& \sum_{i=q,\bar{q}} e_i^2\ f_i(x,\mu_F) \ , \\ 
 A &=& \sum_{i=q,\bar{q}} e_i^2\ A_i(x,\mu_F) \ , \\
 B &=& \sum_{i=q,\bar{q}} e_i^2\ B_i^{{\overline{\mboxsc{MS}}}}(x,\mu_F) \ .
\end{eqnarray}
Inserting the $\sminn$ function (\ref{master}) into Eqn.\
(\ref{2parton}) as a lower integration boundary for each phase-space
point $(x,Q^2)$ will give the complete answer for the total cross
section at NLO. It is important to note that the $\sminn$ function
depends on the factorization and renormalization scales, so that the
improved scale dependence of the NLO cross section as opposed to the
LO one is preserved. As was studied in detail in \cite{4}, the
$\sminn$ function Eqn.~(\ref{master}) is small enough for the soft
and collinear approximations to remain valid which are made to
evaluate the  ${\cal O}(\alpha_s)$  one-parton final states. The NLO
single-jet cross section calculated within the standard approach could
be reproduced within 1--2 \% with this method \cite{4}.

For the calculation of the cross sections in the next section, we have
implemented the function $\sminn$ into the {\tt DISENT} program
\cite{disent} and combined the NLO cross section thus obtained with
the PS algorithm from {\tt PYHTIA} \cite{pythia}. The reason we chose
{\tt DISENT} is to be able to easily extend the results obtained here
to the dijet case later on. In the dijet case also the subtraction
terms from the subtraction method have to be used (see \cite{4} for
details). The steps we have performed to generate events are the
following: 
\begin{itemize} 
\item Define the soft (PS) region by $s_{ij} < R_{tech}Q^2$ for at
  least one pair of partons $i,j$. 

\item The hard region given by $s_{ij}>R_{tech}Q^2$ for all partons $i,j$
  is described by the
  hard two-parton matrix elements (BGF and QCDC). We do not attach the
  PS to the partons of the hard region, although this could in
  principle be done. 

\item For each $(x,Q^2)$ calculate $\sminn$ according to Eqn.\
  (\ref{master}). When $s_{ij}< \sminn$ for any one pair of partons
  $(i,j)$, reject the event. When $s_{ij}>\sminn$ for all $i,j$, 
  start the PS algorithm with the four-vector of the QPM quark as
  input. The weight associated with this
  event is the one obtained from the two-parton final state.
  These weights are positive by construction. 

\item Reject any partons from the PS that lie outside the PS region.
\end{itemize}
The parameter $R_{tech}$ should be larger than $\sminn/Q^2$ but not
too large to ensure that the hard region is appropriately described by
the fixed order matrix elements. $R_{tech}$ is typically of the order
of $1$. We have checked that our cross sections are not sensitive to
the exact value of $R_{tech}$ by varying it by a factor of 2. The
starting scale for the PS is set by the matrix element cutoff
\cite{lepto}.

We have not yet implemented the initital state PS, but
only the final state PS. The parton emission due to the initial state
PS will populate mainly the forward region, which is excluded in the
data with which we are comparing in the next section through
kinematical cuts. Therefore we believe that the effects of the initial
state PS are small. The evolution of the incoming parton is taken into
account through the evolution equations for the partons in the proton.
Furthermore, hadronization effects have not been taken into account
in our calculations. It is however feasible in our model to implement
hadronization, since it can be naturally attached to the PS.

For all calculations in the next section we have 
set the renormalization and factorization scales equal to $Q^2$ and 
employed the CTEQ4M PDFs for the proton.

\section{Results}

The ZEUS collaboration has measured inclusive jet cross sections in
neutral current (NC) DIS $eP$ scattering at $\sqrt{s}\simeq 300$~GeV
for $Q^2>125$~GeV$^2$ \cite{zeus1,zeus1a}. The phase-space of the electron
has been further reduced by restricting its energy $E_{e'}$ and the
inelasticity $y_e$ to $E_{e'}>10$~GeV and $y_e<0.95$, respectively.
The jets were reconstructed using a $k_T$ cluster algorithm \cite{kt}
in the laboratory frame. The reconstructed jets were required to have
a minimum transverse energy of $E_T>14$~GeV and a pseudorapidity in
the range $-1<\eta<2$. The data are corrected for detector effects.

In Fig.~\ref{fig-et} we show the comparison of our calculation, given by
the full line labeled '{\tt DISSET}', to the ZEUS data in four regions
of $Q^2$, namely $Q^2>125, 500, 1000$ and $2000$~GeV$^2$. We see an overall
good agreement. Since our predictions are on parton level, we have
checked the influence of the hadronization using the {\tt LEPTO} event
generator \cite{lepto}. We found that the overall effect of the
hadronization was small, below 5\%, with a tendency to lower the
parton level predictions. We have furthermore checked that the PS
changes the $E_T$ distribution only marginally, also below 5\%. In
\cite{zeus1a} the data were compared to a standard NLO calculation with
{\tt DISENT} \cite{disent}. We have redone these calculations for
comparison and also plotted the results in Fig.~\ref{fig-et}, shown as
the dashed line labeled 'standard {\tt DISENT}'. Our results are in
agreement with those from \cite{zeus1a}. The agreement between data
and NLO theory is similarly good as for our calculation. From this
comparison we deduce that our calculation reproduces correctly the
standard NLO result, which confirms the findings in \cite{4}. 
We emphasize that the correct normalization of the cross sections is a
non-trivial result of our procedure. We obtained the normalization
without using information from the total $eP$ scattering cross
section. The normalization comes directly from the NLO matrix
elements, modified by the $\sminn$ function.

A similarly good description of the data is seen in Fig.~\ref{fig-eta} for
the $\eta$ distribution, which is integrated over $E_T>14$~GeV and 
$Q^2>125$~GeV$^2$. The data are described rather well, both with the
standard NLO calculation as well as with our new calculation. 

We proceed to a comparison of our calculation with jet shape
measurements. The ZEUS collaboration has measured the differential and
integrated shape of jets in neutral current DIS events with
$Q^2>100$~GeV$^2$ for jets with $E_T$ above 14~GeV and $-1<\eta<2$
\cite{zeus2}. The jets are reconstructed using an iterative cone
algorithm in the $(\eta,\phi)$ plane \cite{cone2,snow}. For the
scattered electron further restrictions are $E_{e'}>10$~GeV and
$y_e<0.95$, as for the previously discussed data. The differential jet
shape is defined as the average fraction of the jet's transverse
energy that lies inside an annulus in the $(\eta,\phi)$ plane of inner
(outer) radius $r-\Delta r/2$ ($r+\Delta r/2$) concentric with the jet
defining cone
\cite{sdellis}: 
\begin{equation}
\label{eqjsd}
 \rho(r)=\frac{1}{N_{jets}} \frac{1}{\Delta r}
 \sum_{jets}\frac{E_T(r-\Delta r/2,r+\Delta r/2)}{E_T(0,R)} \ . 
\end{equation}
Here, $E_T(r-\Delta r/2,r+\Delta r/2)$ is the transverse energy within the 
given annulus and $N_{jets}$ is the total number of jets in the sample.
The differential jet shape has been measured for $r$ values varying from 
$0.05$ to $0.95$ in $\Delta r=0.1$ increments. The integrated jet shape 
is defined by
\begin{equation}
\label{eqjsi}
  \Psi(r) = \frac{1}{N_{jets}} \sum_{jets} \frac{E_T(0,r)}{E_T(0,R)} \ .
\end{equation}
By definition, $\Psi(R)=1$. It has been measured for $r$
values varying from $0.1$ to $1.0$ also in increments of $\Delta r=0.1$ 

\begin{figure}[hhh]
  \unitlength1mm
  \begin{picture}(122,160)
    \put(3,0){\epsfig{file=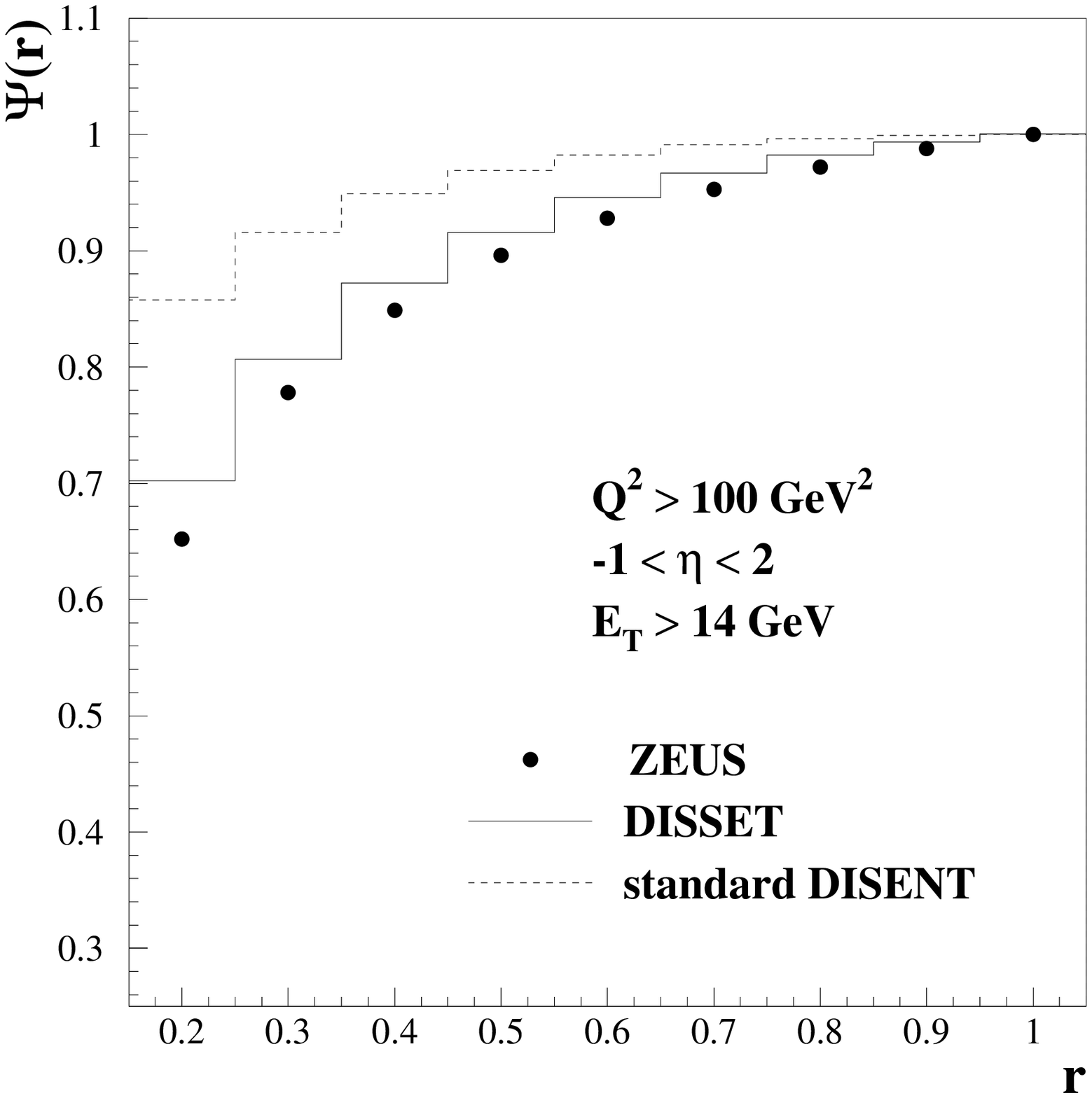,width=16cm}}
  \end{picture}
\caption{Integrated jet shape $\Psi(r)$ in NC DIS events integrated
 over $Q^2 > 100$~GeV$^2$ for jets with $E_T>14$~GeV and
 $-1<\eta<2$. The ZEUS data are compared to our calculation 
 (full line) and to a NLO calculation (dashed line).}
\label{fig-shape}
\end{figure}

In Fig.~\ref{fig-shape} we compare our calculation (full line) and 
a standard NLO calculation (dashed line) to the ZEUS data which are
corrected to hadron level. The description of the jet shape within our
approach is rather good. The jets produced are slightly too narrow,
our calulation being about 8\% above the data in the lowest $r$ bin. We
have checked with {\tt LEPTO} that the hadronization effects can account
for the difference between data and prediction, i.e.\ including
hadronization in our calculation would bring our results to good
agreement with the data.  We furthermore see that the
standard NLO prediction is much too large for all $r$. This implies
that the jets produced by the standard calculation are not broad
enough. Reducing the cross sections by taking into account
hadronization is not sufficient for the standard NLO calculation. This
result is not surprising since the standard NLO calculation provides only a
LO prediction for the jet shape in which at maximum two partons are
combined to give a jet.

\begin{figure}[hhh]
  \unitlength1mm
  \begin{picture}(122,160)
    \put(3,0){\epsfig{file=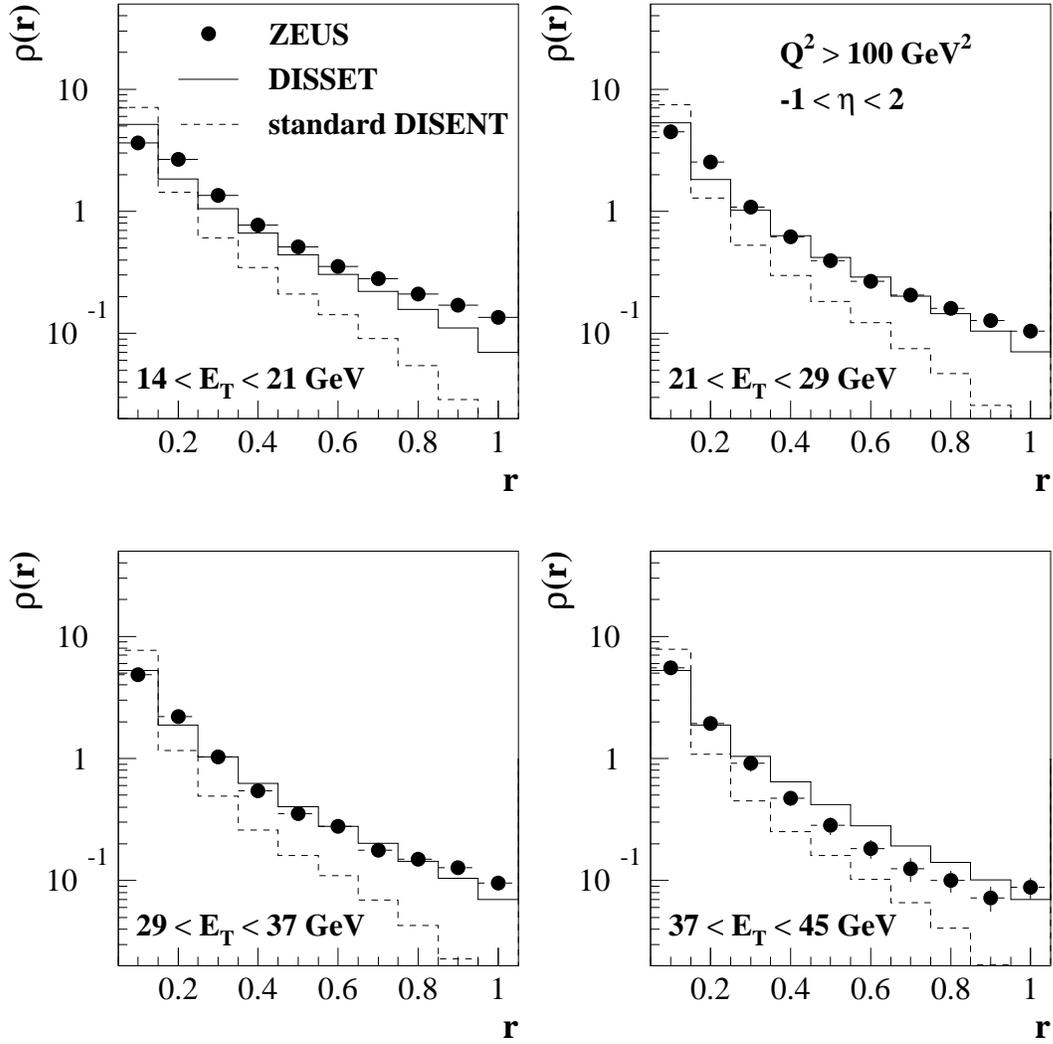,width=16cm}}
  \end{picture}
\caption{Integrated jet shape $\rho(r)$ in NC DIS events integrated
 over $Q^2 > 100$~GeV$^2$ for jets with $-1<\eta<2$ in four different
 ranges of $E_T$. The ZEUS data are compared to our calculation 
 (full line) and to a NLO calculation (dashed line).}
\label{fig-dshape}
\end{figure}

In Fig.~\ref{fig-dshape} we finally compare our calculations to the
ZEUS differential jet shapes for four different $E_T$ regions. We find
similar results as for the integrated jet shapes. The standard NLO
calculation (dashed line) is clearly too narrow in all four $E_T$
regions, whereas our calculation (full line) gives a much  better
description. In the lowest $E_T$ bin our calculation produces slightly
too narrow jets, whereas in the largest $E_T$ bin the jets tend to be
too broad. Using {\tt LEPTO}, we have checked the hadronization
corrections also for the differential jet shapes and found the
discrepancies between our calculation on the parton level and the data
can be accounted for by hadronization effects.

\section{Summary and Outlook}

We have investigated the method described in \cite{4} to combine a
fixed NLO QCD calculation, including the real soft and collinear
as well as the virtual corrections, with a leading log PS. We have 
implemented the method in {\tt DISENT} and attached the PS from 
{\tt PYTHIA}. We have compared our calculation with data from ZEUS, both
for inclusive jet cross sections in the laboratory frame, as well as
for jet shape measurements. For both measurements, good agreement with
our model was found. In addition, good agreement was found between the
standard NLO calculation and our results for the case of the inclusive
spectra in transverse energy and rapidity, where the influence of the
PS is marginal. This shows that we have obtained the correct NLO
normalization of the cross sections together with a good description
of the internal structure of the jets.

To obtain a full event generator, the initial state PS as well as
hadronization need to be implemented in our program package. This will
be done in the near future. The next, more complicated step is to
proceed to the two-jet final states and combine the NLO calculations
for this case with the PS. Although some details have to be clarified
for this case, we could show the principle feasibility of our method.

\bigskip

\subsection*{Acknowledgements}

\begin{figure}[hhh]
  \unitlength1mm
  \begin{picture}(122,160)
    \put(3,0){\epsfig{file=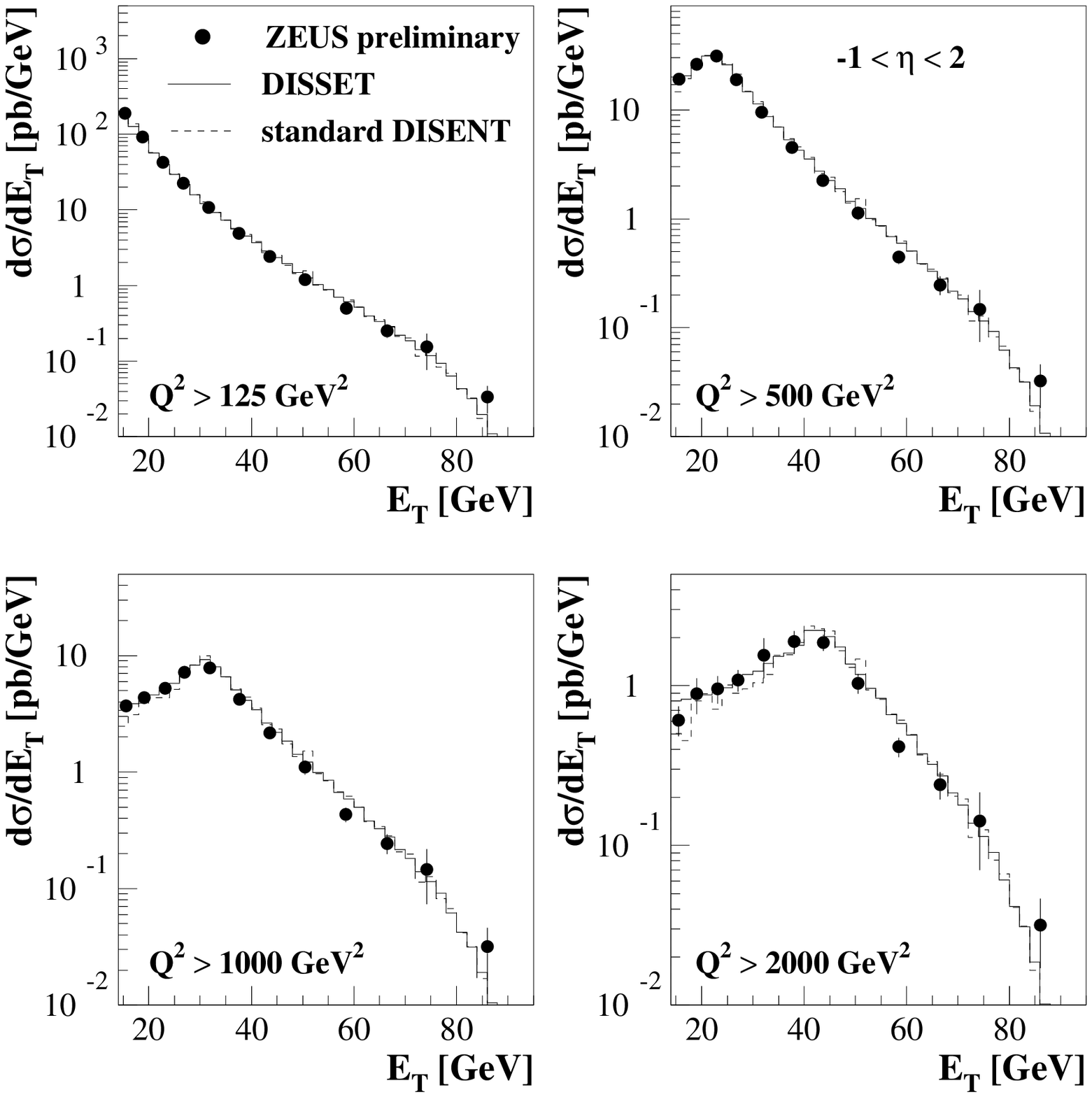,width=16cm}}
  \end{picture}
\caption{Differential cross sections $d\sigma/dE_T$ for inclusive jet
production in NC DIS events integrated over $-1<\eta<2$ for
$Q^2>125,500,1000$ and $2000$~GeV$^2$. The ZEUS data are compared to our
calculation labeled '{\tt DISSET}' (full line) and to a NLO
calculation, labeled 'standard {\tt DISENT}' (dashed line).}
\label{fig-et}
\end{figure}

\begin{figure}[hhh]
  \unitlength1mm
  \begin{picture}(122,160)
    \put(3,0){\epsfig{file=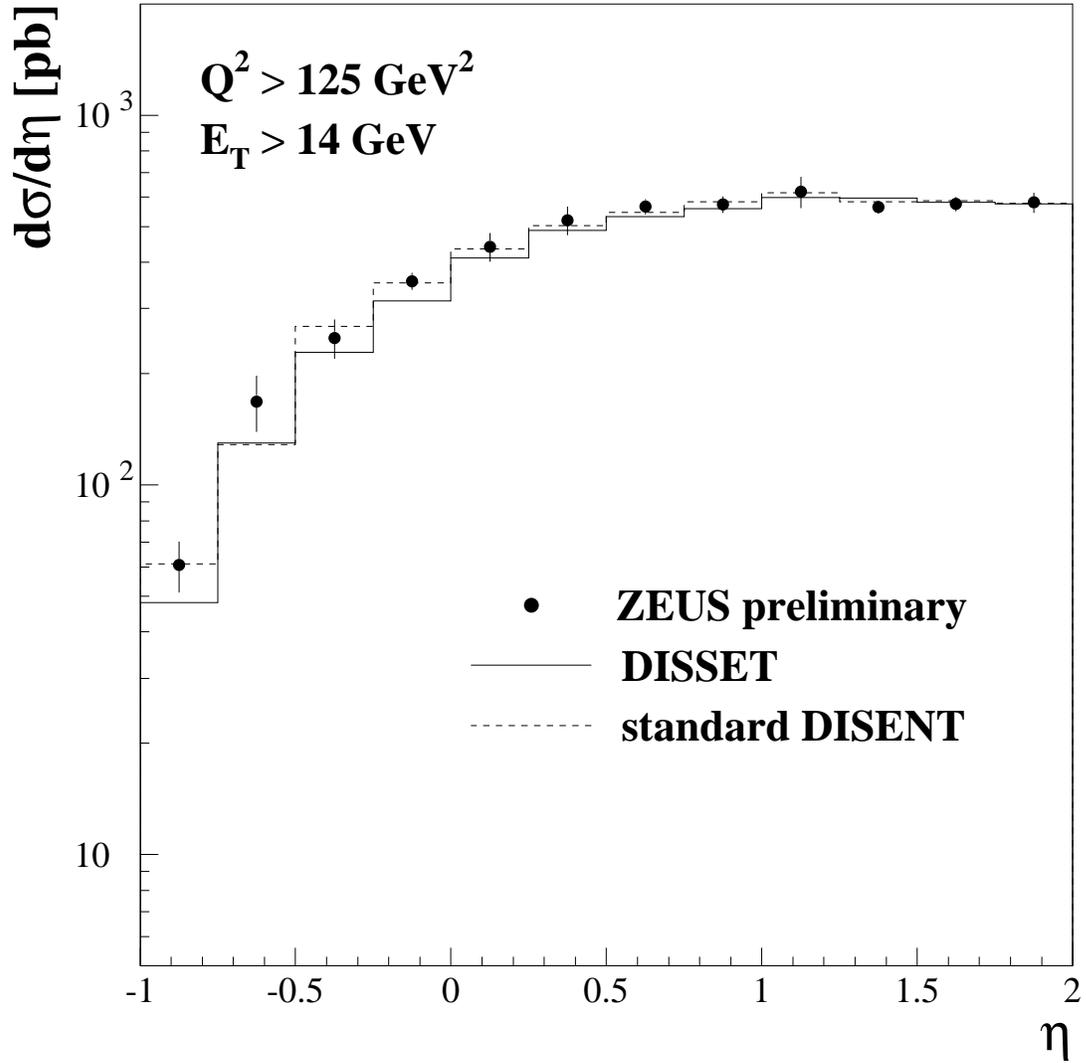,width=16cm}}
  \end{picture}
\caption{Differential cross sections $d\sigma/d\eta$ for inclusive jet
production in NC DIS events integrated over $E_T>14$~GeV and
$Q^2>125$~GeV$^2$. The ZEUS data are compared to our calculation (full line)
and to a NLO calculation (dashed line).}
\label{fig-eta}
\end{figure}

We are grateful to J.~Terron for providing us with the data points of 
Figs.~\ref{fig-et} and \ref{fig-eta}.

\newpage

\end{document}